\documentclass{article}
\usepackage{spconf,amsmath,graphicx,hyperref,cite}

\def\x{{\mathbf x}}

\title{A NEW FRAMEWORK FOR SPATIAL MODELING AND SYNTHESIS OF GENOME SEQUENCE}
\name{Salman Mohamadi, Farhang Yeganegi, Hamidreza Amindavar}
\address{Department of Electrical Engineering, Amirkabir University of Technology, Tehran, Iran\\
	\tt S.mohamadi100@aut.ac.ir, Farhang.yeganegi@aut.ac.ir, Hamidami@aut.ac.ir}
\begin{document}
%
\maketitle
\begin{abstract}
This paper provides a framework in order to statistically model sequences from human genome, which is allowing a formulation to synthesize gene sequences. We start by converting the alphabetic sequence of genome to decimal sequence by Huffman coding. Then, this decimal sequence is decomposed by HP filter into two components, trend and cyclic. Next, a statistical modeling, ARIMA-GARCH, is implemented on trend component exhibiting heteroskedasticity, autoregressive integrated moving average (ARIMA) to capture the linear characteristics of the sequence and later, generalized autoregressive conditional heteroskedasticity (GARCH) is then appropriated for the statistical nonlinearity of genome sequence.
This modeling approach synthesizes a given genome sequence regarding to its statistical features. Finally, the PDF of a given sequence is estimated using Gaussian mixture model, and based on estimated PDF, we determine a new PDF presenting sequences that counteract statistically the original sequence. 
Our strategy is performed on several genes as well as HIV nocleotide sequence and corresponding results is presented.
\end{abstract}
\section{Introduction}
\label{sec:intro}

Considering the increasingly enormous amount of human and also animals genomic data and assuming that genome sequences could also be treated as a discrete signal, raise the idea that signal processing can play an important role in interpreting the genome information \cite{mabrouk2017different,vaidyanathan2004role}.
Here we analyze and model genome sequence which is represented by sequence of letters from a four-character alphabet including of the letters A, T, C, and G; each one representing one of four distinct nucleotides named cytosine (C), thymine (T), adenine (A), guanine (G). Also A, T, G and C are called base pairs of gene sequence. DNA strand is divided into genes and intergenic subsequences noting that genes are responsible for protein synthesis. Therefore in order to process a sub sequence of DNA, here a sequence demonstrating a gene, we need to convert the sequence of alphabet to a sequence of binary or decimal numbers \cite{das2017novel}. A variety of methods have been used to convert the sequence of alphabet to desirable decimal sequence which can be found in \cite{vaidyanathan2004role,das2017novel,naseem2017preprocessing}. Sequence of any particular gene, is a reference to possessing necessary information in order to create a specific protein; so any time that a new protein is going to come into existence, a translation system start to translate the gene information letter by letter. In this sense, we can suppose that we are facing a microscopic communication system that to transfer information. Therefore in the procedure of translating genome sequence, it can be infered that genome sequence is the source of information transmission during genome sequence biological translation \cite{bossert2017information,thangavel2017comparative} and based on this fact, we suggest a source coding method for converting the gene sequence to a decimal sequence. After providing a digital signal for any given gene of genome sequence, since frequently have been observed that genome sequence demonstrate cyclic pattern \cite{shabalina2006periodic}, a powerful tools named HP filter will be applied to this decimal sequence to decompose it to trend component and Cyclic component. Cyclic present periodic patterns of signal, but trend provide the main trend of signal. Also it is important to note that cyclic component shows a time invariant variance but trend component shows an entirely time variant variance, it can be seen in Fig.~\ref{fig2}. Finally based on ARIMA-GARCH model, modeling and synthesis of genome sequences, is done for a number of original genes sequences. This paper is organized as follows, the preprocessing of the genome sequence, ARIMA and ARCH/GARCH modeling of the genome sequence, PDF estimation and some simulations and at the end, some concluding remarks.
\section{Problem Statement and Formulation}
In order for statistical modeling of a genome sequence on a computer, decimal numbers are preferred, on the other hand, Huffman coding is a variable-length coding commonly accepted for lossless data compression providing a short binary sequence; in entropy sense, easily converted to a decimal number. By so doing, a discrete signal like a time series is created of a genome sequence that measures the status of genetic alphabets over time and space which later would be used for spatial modeling. There is no evidence as to the created sequence is stationary, in order to convert this non-stationary series to stationary, in fact, having a look into the genome structure, because genome sequence includes too many sub sequences such as sequences of genes which are randomly arranged from statistical point of view \cite{zhao2017genome}; ARIMA modeling is utilized that captures the best linear forecast for the genome sequence. In order to model, the recipe of the frequency of how the alphabets in a genome sequence converted to each other a statistical measure of the dispersion can be adopted, volatility is the measure of time-varying variance; which is the first order non-linearity characteristics of a time series, and it has been modeled through ARCH/GARCH method. As we demonstrate that the trend component of genome sequence possesses volatility clustering reflecting more recent changes and fluctuations in the genome time series.

\subsection{PREPROCESSING}
\subsubsection{Huffman coding}
There are two types of coding algorithm in terms of code length which are fixed length code and variable length code. Along with all the advantageous that variable length coding has, they may not be uniquely decodable. In fact, in coding theory, a specific type of code called "prefix code" can be uniquely decoded. The other aspect of coding algorithm that is really important is the average code length. In information theory litretures, it can be seen that the average code length has a lower bound called "Entropy". Also, this bound can be achieved using a prefix variable length code called "Huffman code" where it has shown a promising results in the litretures \cite{rajarajeswari2011dnabit,yeganegi2018comparative}. The output from Huffman's algorithm can be viewed as a variable-length code table for encoding a source symbol. The algorithm derives this table from the estimated probability or frequency of occurrence (weight) for each possible value of the source symbol. In other words, in this method, probability of symbols (here A, T, G and C) are take into account to prepare a code specifically efficient for the alphabetic sequence we want to encode it \cite{he2017practice}. As in \cite{al2017toward} have been demonstrated, Huffman coding would serve as good compression code for genome sequence, virtually because of presenting sequence-specific or here, gene-specific coding, meaning that with this method, entropy of each of the nucleotides, A, T, G and C, individually affect the coding procedure of that sequence. Therefore by this method, the task of converting the sequence of A, T, G and C, to a digital signal would be performed efficiently \cite{al2017toward}. With Huffman method, we map [A T G C] to decimal numbers in vector N = [0 1 2 3], noting it has been seen that different mappings, result in the same ways, so we can map [A T G C] to any other vector made up from 0, 1, 2 and 3. \\ 

\subsubsection{Hodrick-–Prescott (HP) filtering }
 
\begin{figure}[t]
	\centering{
		\includegraphics[scale=.12]{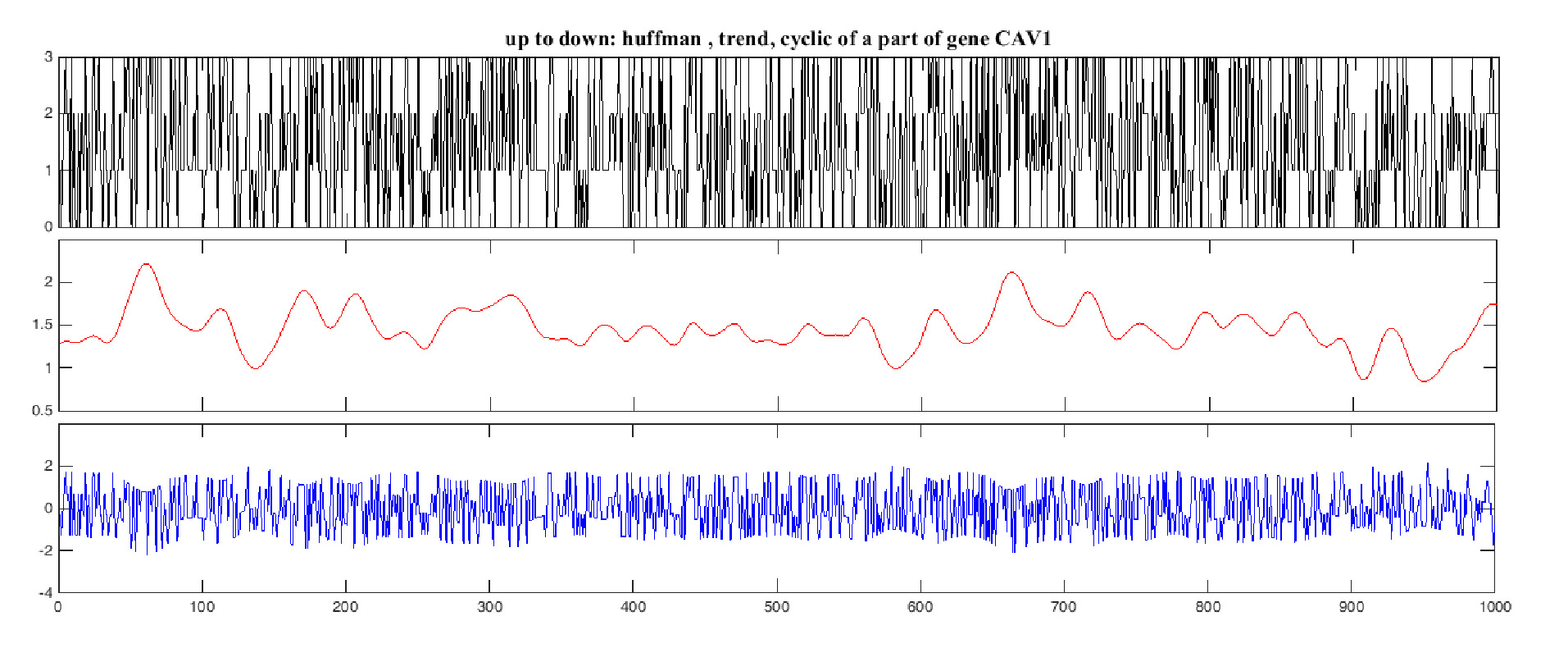}
		\caption{Huffman encoded decimal sequence of 1000 base pairs (letter) of gene CAV1, trend Component  and Cyclic Component of this decimal sequence, horizontal axis represents base pair and veritacl axis represents amplitude .}
		\label{m_estimate}}
\end{figure}

\begin{figure}[t]
			\centering{
				\includegraphics[scale=.65]{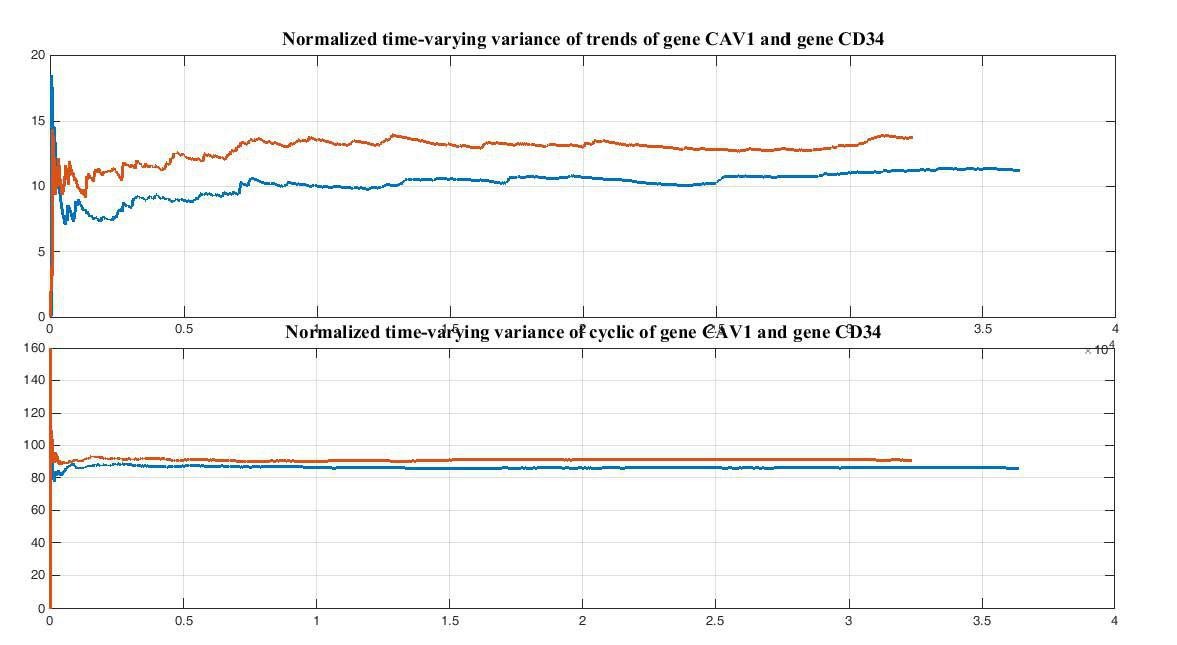}
				\caption{Normalized Time Varying Variance of Trend and Cyclic components of the two genes, CAV1 (blue) and CD34 (orange); horizontal axis represents base pairs (unit*10000) and veritacl axis represents amplitude.}
				\label{fig2}}
		\end{figure}

Hodrick–-Prescott filter (HP filter) captures a smooth trend carrying the most valuable information of signals rather than powerful hidden cyclic pattern \cite{pedersen2001hodrick,bruchez2003modification}. The HP filter is a fine approximation to an ideal high-pass filter, which outputs two components named cyclic and trend. The genome sequence converted to a decimal sequence can be decomposed into two components: a trend and cycle components, and a remainder component that may not be captured denoted as noise. This is a problem of interest in the statistical modeling, extraction of trend and hidden periodicity from a time series which may seem not to carry valuable information. However this question remains unaddressed that why we should use this decomposition. Cyclic component is about time invarying variance part of signal, therefore once we aim at modeling the signal with a model capturing time varying variance, we may decompose the signal into time variant and time invariant components and just model the first component. In section 4 we explain in more detail how we use formulation of our model and also intact cyclic component of certain gene to synthesize it. 
The trend is the component of a genome sequence that provides the low frequency statistical characteristics, whereas the cyclical component that refers to the fluctuations around the trend. In a genome sequence, the cyclic component does not convey a significant information, in entropy sense. In Fig.~\ref{m_estimate} Huffman encoded sequences of cyclic and trend are shown for a limited number of base pairs (letters of alphabetic sequence) of gene. Furthermore, their time varying variance to support the idea that HP filter is effective in this application, additional to above argument, it is helpful to consider that trend component presents that component of signal with fluctuating time varying variance, meaning that trend possesses the heteroskedasticity while cyclic component presents those of almost constant time invarying variance, meaning cyclic is a homoskedastic process; as it is shown in Fig.~\ref{fig2}. Heteroscedasticity characteristics of genome sequence refers to the circumstance in which the variability of the trend component cannot be predicted  by any predictor if equal ranges of genome sequence are observed, on the other hand, a prediction of genome sequence is possible of if unequal ranges of trend values are observed.
\subsection{ARIMA--ARCH/GARCH modeling}
\subsubsection{ARIMA Model}
\label{sec:typestyle}
A genome sequence possesses some properties that exist among all other genes in a linear sense, from the standpoint of forecasting these properties ARIMA modeling allows linear prediction of future values of genome sequence based on these properties.
\label{sec:typestyle}
ARIMA models are applied in cases that data shows non-stationarity, in order to reduce non-stationarity. Via applying an initial differencing step, corresponding to the integrated part of the model, ARIMA reduces the non-stationarity \cite{bollerslev1986generalized}. The ARIMA( ${p}$;  ${d}$;  ${q}$) is described by 
\begin{equation}
	\label{eq01}
	\left( {1 - \sum\limits_{k = 1}^p {{a_k}} {L^k}} \right){\left( {1 - L} \right)^d}{X_n} = \left( {1 + \sum\limits_{k = 1}^q {{b_k}} {L^k}} \right){\epsilon _n}
\end{equation}
where ${X_n}$, here is the decimal sequence corresponding to gene sequence, ${n}$ is the number corresponding to the location of base pairs in the sequence of gene, for example ${X_{128}}$ refers to the decimal number in the 128th location of gene that corresponds to the type of base pair that has occupied this location of gene sequence.\\
${ACCCACATTCCCCTCTCCA...''128th base pair''...}$\\
Moreover ${L}$ is the lag operator, the ${a_k}$ are the parameters of the autoregressive part of the model, the ${b_k}$ are the parameters of the moving average part and the ${\epsilon _n}$ are error terms. The error terms ${\epsilon _n}$ are generally assumed to be independent, identically distributed variables sampled from a normal distribution with zero mean. Here ${p}$ is the order of AR, and ${q}$ the order of MA, and ${d}$ is the order of differencing \cite{mohamadi2017arima}.

\renewcommand{\arraystretch}{1.3}
\newcommand{\head}[1]{\textnormal{\textbf{#1}}}
\begin{table}[h]
	\centering{
		\caption{Parameters of the ARIMA models;
			$N$ is the number of base pairs (letter) of sequence of gene CAV1
			.}
		\label{table_parameter1}
		\begin{tabular}{cc||cc}
			\hline
			\head{Model} & \head{N} &  \head{AIC$_{\mbox{\small c}}$} \\
			\hline
			\hline
			
			ARIMA(1,1,1)           & 38400                       & -5.1563e+05               \\
			ARIMA(2,1,1)                 & 38400                 & -5.1568e+05  \\
			ARIMA(2,1,2)               & 38400                  & -5.1616e+05                \\
			ARIMA(1,1,2)                  & 38400                   & -5.1533e+05               \\
			ARIMA(2,2,2)                       & 38400                 & -5.1601e+05                \\
			ARIMA(2,2,1)               & 38400                  & -5.1586e+05               \\
			ARIMA(1,2,2)                   & 38400               & -5.1583e+05              \\
			ARIMA(1,2,1)   & 38400     &     -5.1558e+05             \\
			
		\end{tabular}}
	\end{table}
	
	
	\begin{table}[h]
		\centering{
			\caption{Parameters of the ARCH and GARCH models; N also is the length of residual of ARIMA model which is equal to the size of gene CAV1, here we see GARCH(1,1) is the best of all due to  the smallest
				AIC$_{\mbox{\small c}}$.}
			\label{table_parameter}
			\begin{tabular}{cc||cc}
				\hline
				\head{Model} & \head{N} & \head{AIC$_{\mbox{\small c}}$} \\
				\hline
				\hline
				ARCH(0)           & 38400                       &  -2.889e+05           \\
				ARCH(1)           & 38400                       &  -2.8913e+05           \\
				ARCH(2)           & 38400                       &   -2.9221e+05           \\ 	
				ARCH(3)           & 38400                       &  -2.9614e+05                       \\
				ARCH(4)                 & 38400                  & -2.9905e+05          \\
				ARCH(5)               & 38400                 &  -3.1421e+05           \\
				ARCH(6)         &  38400             & -3.2228e+05              \\
				ARCH(7)                       & 38400                  &  -3.3633e+05                               \\
				ARCH(8)               & 38400                &   -3.4689e+05             \\
				ARCH(9)                   & 38400                    &  -3.5656e+05                          \\
				GARCH(1,1)   & 38400    &   -3.9555e+05               \\

			\end{tabular}}
		\end{table}
\subsubsection{ARCH or GARCH Modeling}
Most straight forward of time varying power of random sequences is ARCH/GARCH; it means that these processes best model the sequences of time varying power or variance. ARCH/GARCH process mostly and also first time, used in economic time series modeling, because its profound mathematics allow to cover precisely most of fluctuations in those signals. Here because of lack of space, we just present a brief introduction, for further information and also example of biological application of these two, refer to \cite{bollerslev1986generalized,mohamadi2017arima} [15,14]. For a given digital signal ${y_n}$, we can model it using GARCH( ${p}$, ${q}$), if ${E \big\{y_n\big\} = 0}$ and 
		\begin{eqnarray}
			\label{equ1}
			{y_n} = \sqrt {{h_n}} {\epsilon_n} \\
			{h_n} = {a_0} + \sum\limits_{i = 1}^p {{a_i}} {y^2}_{n - i} + \sum\limits_{j = 1}^q {{b_j}} {h_{n - j}}
		\end{eqnarray}
		Where ${h_n}$ is the conditional variance of ${{y_n}}$, ${a_0 > 0}$, ${a_i  \geq  0}$, ${b_j  \geq  0}$ and ${\{\epsilon_n}\}$ is a sequence of independent identically distributed zero mean with variance one random variables, often assumed to be a standard normal random variable \cite{mohamadi2017arima}. ARCH or GARCH process, with an ability to model volatility, would be used to model the volatility of the digital signal to reflect changes and fluctuations in gene sequence \cite{mohamadi2019detection}.
\section{FITTING THE BEST ARIMA-GARCH MODEL}
		To perform our method, we selected 18 genes that have most interaction in human breast cancer; the sequence of each gene is downloaded from United State National Center for Biotechnology Information\footnote{https://www.ncbi.nlm.nih.gov/gene/}. These genes are as in table 3. Having  the preprocessed digital signal, corresponding to each gene sequence, now different ARIMA(p; d; q) models are fitted and the corresponding AICc are computed as in Table 1. Best ARIMA for sample gene, CAV1 is ARIMA(2,1,2). When it comes to fit a ARCH/GARCH model, firstly ARCH test should be used to verify the existence of heteroskedasticity in the data. For all 18 genes this test has been performed on residual of ARIMA model individually and confirmed the issue. After that, best orders are extracted through AIC. Finally a mathematical formulation resulting from modeling is attained.
		
\section{Results}
\subsection{Modeling the sequences of 18 genes of human genome}
Order of ARIMA model for each gene is selected based on best AIC. Best ARCH/GARCH model, is then fitted to residuals of ARIMA. Results of sample gene, gene CAV1, are presented in tables and figures. For all of genes we achieved mathematical formulation extracted from ARIMA-GARCH modeling; here 15 out of 18 genes are modeled by formulation of ARIMA(2,1,2)-GARCH(1,1) model, only the coefficients of formula differs from gene to gene. Among 18 genes, except for three genes PPARG, NGFR and PLA2G2A, best ARIMA was ARIMA(2,1,2) and for all 18 genes, GARCH(1,1) was selected as the best order of GARCH. For these three, PPARG, NGFR and PLA2G2, best ARIMAs were respectively ARIMA(2,2,2), ARIMA(2,2,2) and ARIMA(1,2,2). Fitting these best models, allow us to synthesize artificial gene sequence for any given original gene, that possess the linear and nonlinear characteristics of the original gene.

\subsection{Synthesizing the new sequences}
According to the formulation provided at 2.2 and 2.3; we can formulate ARIMA-GARCH model \cite{mohamadi2017arima}. For all genes, coefficients of the formula of ARIMA-GARCH modeling are extracted; here for gene CAV1 with ARIMA(2,1,2)-GARCH(1,1) we have: 
\begin{eqnarray*}
	h_n+X_n-X_{n-1}&=&\mbox{($-0.4369$)}(X_{n-1}-X_{n-2})\\
	&&+\mbox{($0.1312$)}(X_{n-2}-X_{n-3})\\
	&&+\mbox{($-0.1334$)}\epsilon_{n-1}+\mbox{($-0.8718$)}\epsilon_{n-2}\\
	&&+\mbox{($0.8991$)}\epsilon^2_{n-1}+\mbox{($0.0189$)}h_{n-1}\\
	&&+\mbox{($7.40609e-06$)}
\end{eqnarray*}
Where ${h_n}$ is synthesized sequence of Trend component of the artificial sequence, which is going to be synthesized completely, and ${X_n}$ is the sequence of Trend component of the original gene and ${\epsilon_n}$ the error term. Then we add ${h_n}$ to the cyclic component of the original gene to complete the synthesis procedure, which provides us a new decimal sequence for any given original sequence. The latter, using Huffman decoding, we convert the synthesized sequence, that is decimal, to alphabetic sequence.\\
${ \big\{d_1,d_2,d_3...d_m\big\}   \Rightarrow  \big\{b_1,b_2,b_3,...b_m\big\} }$\\
${d_1,d_2,...d_m}$ are the decimal numbers resulting from synthesis and ${b_1,b_2,b_3...b_m}$ are the corresponding base pairs of the synthesized gene which are achieved by converting the decimal sequence to alphabetic sequence. Sample of 120 synthesized base pairs (from 1332 to 1452), which is randomly selected from whole sequence of synthesized gene base on gene FGF2 are as follow :\\
ACGCCCGCGAGGCTGGTTGGCCGGGGCGGGGGCCA\\
TGCCGCGGAGCGTGTCGGCGGCCGAGGCCGGCGCC\\
GTAGGACGGCGGCTCAAAGCGCGGCTCCATCGACTC\\
GCAGATCTCGGGGG\\
It is noteworthy to mention that in the sense of practical application, the above formulation equips us so that we can synthesize new sequences based on a certain genomic sequence, which is useful in some application such as genomic sequence PDF estimation, because for PDF estimation we need to provide enough ensemble of the genomic sequence, which this formulation can deal with.
\begin{table}[h]
	\centering{
		\caption{List of 18 new sequences, which  are synthesized using Huffman decoding based on original genes; 
			$N $ is the length of the new sequence which is the same length as  of original gene}
		\label{table_parameter1}
		\begin{tabular}{cc||cc}
			\hline
			\head{Original gene} & \head{$N$} &  \	\head{Original gene} & \head{$N$} \\
			\hline
			\hline
			
			IGFBP6           & 4910                       & TGFBR3   & 225993           \\
			NGFR                 & 19728                 & EGR1  & 3826  \\
			ESR1               & 472929                  & FOS & 3461                \\
			DLC1                  & 521250                   & IGF1 & 85920               \\
			CD34                       & 30431                 & NTRK2 & 358613               \\
			CAV1               & 38401                 & PLA2G2A & 5009               \\
			PPARG                   & 183646               & ERBB4 & 1163439             \\
			FGF2   & 71529    &     IL6 & 6561             \\
			
		\end{tabular}}
	\end{table}
	
		\begin{figure}[t]
			\centering{
				\includegraphics[scale=.13]{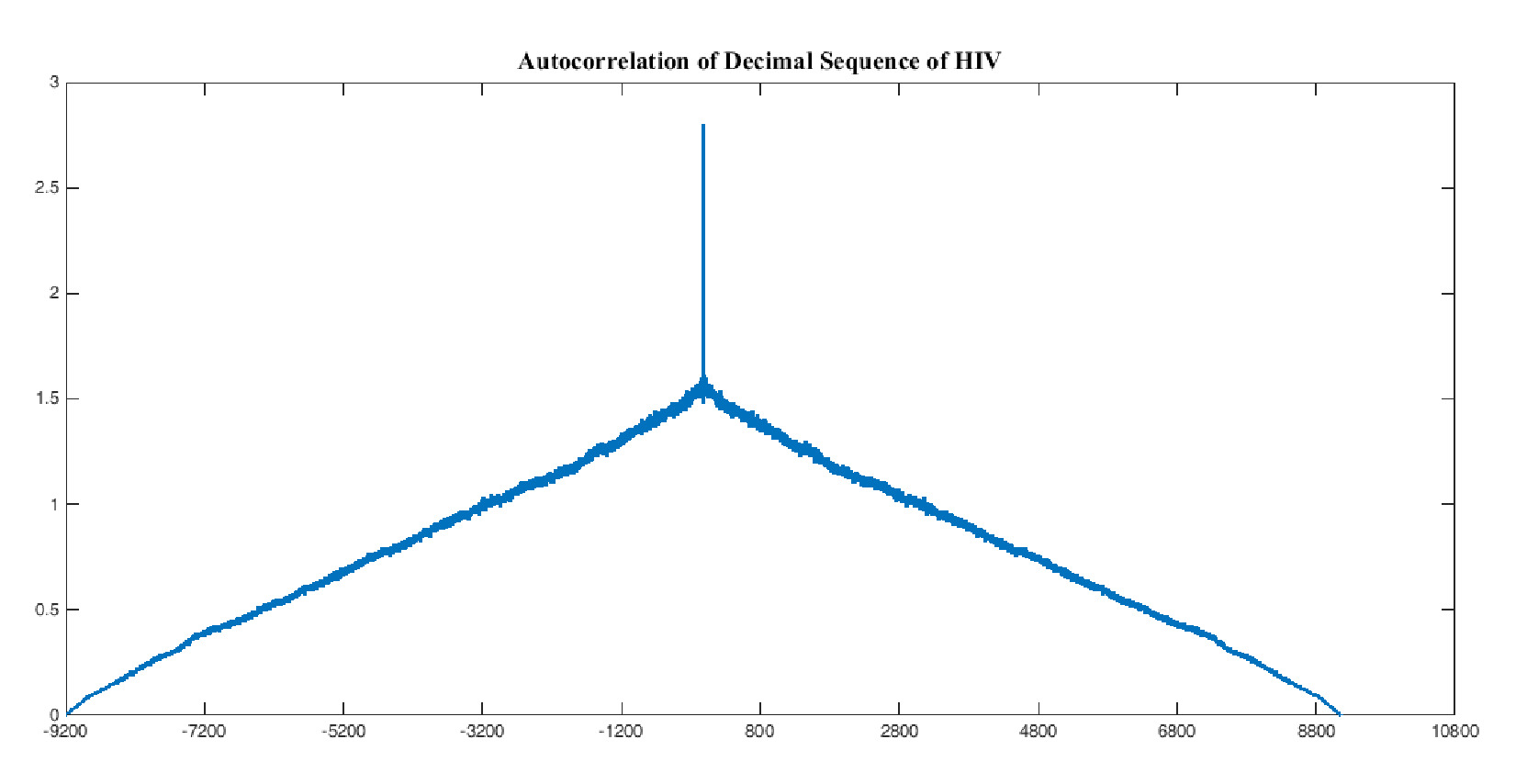}
				\caption{HIV Sequence Autocorrelation, as we see, even at heavy lags, the sequence is highly correlated.}
				\label{autocorr}}
		\end{figure}

		\begin{figure}[t]
			\centering{
				\includegraphics[scale=.098]{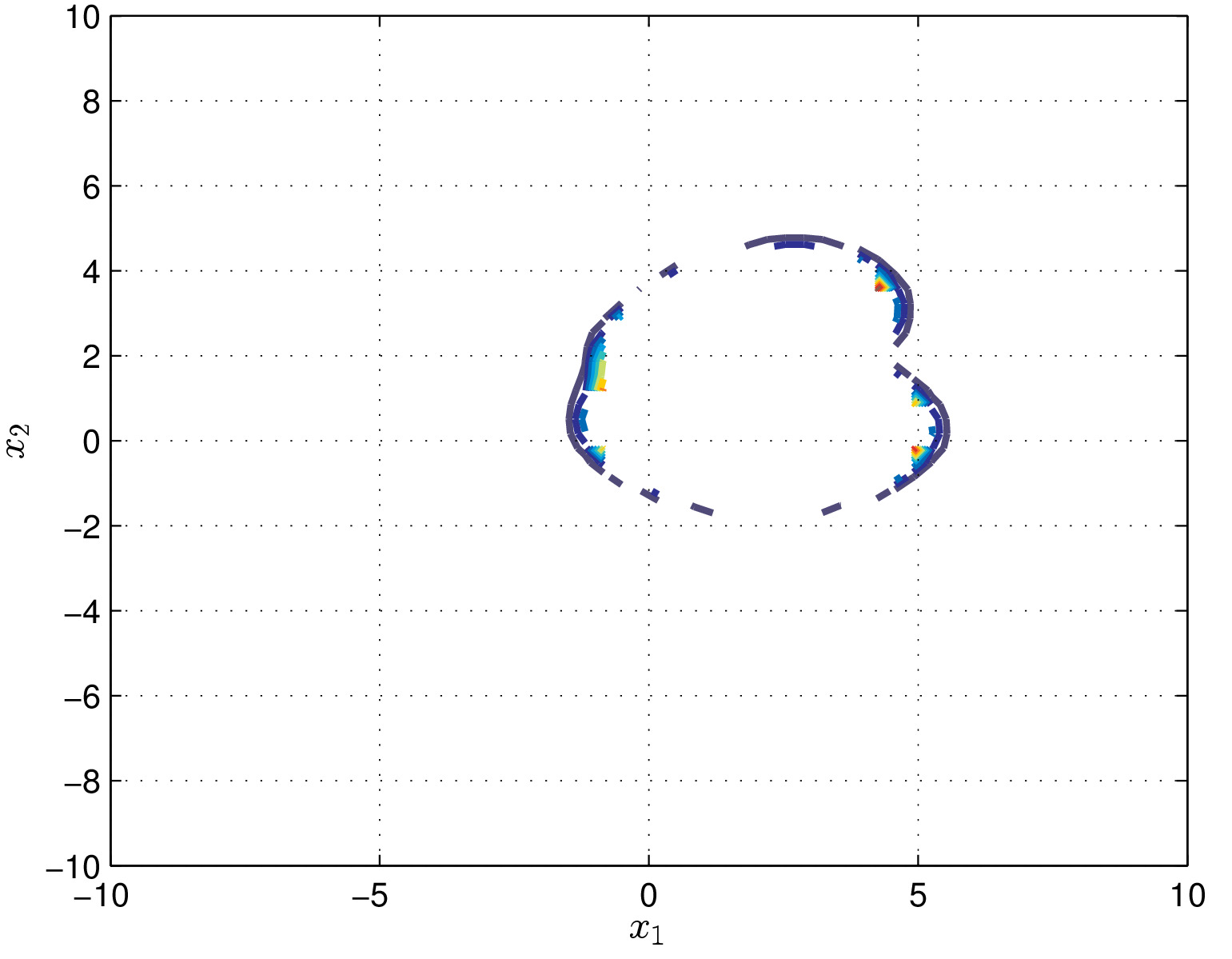}
				\caption{Gaussian mixture approximation to PDF of HIV sequence.}
				\label{PDF1}}
		\end{figure}

		\begin{figure}[t]
			\centering{
				\includegraphics[scale=.4]{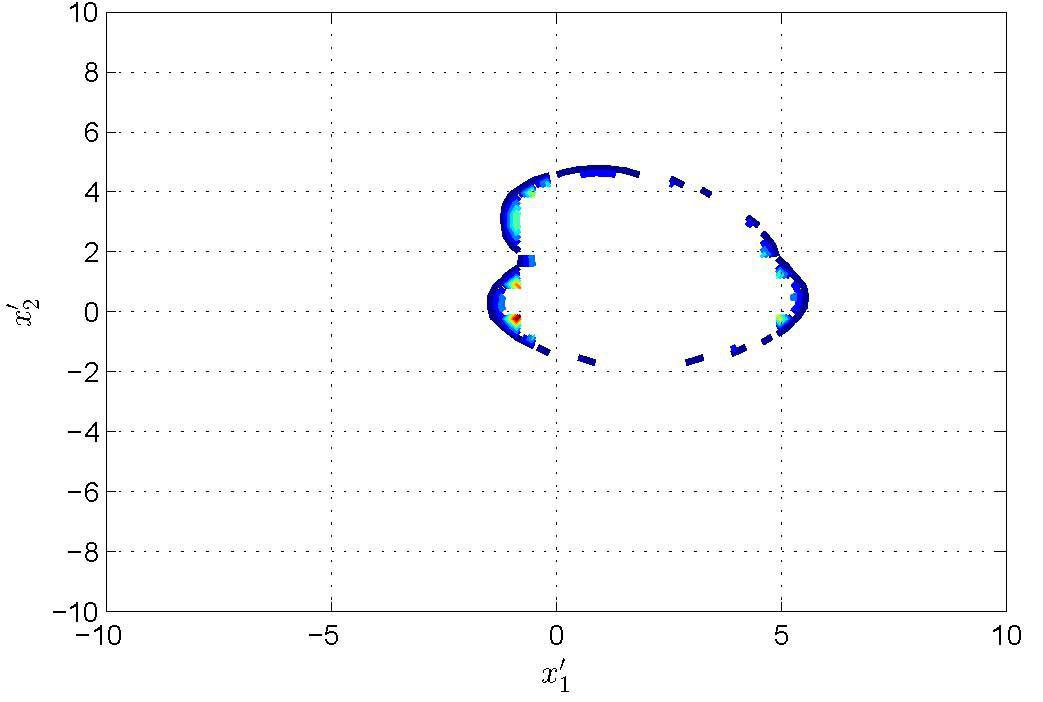}
				\caption{Gaussian mixture approximation to PDF of highly uncorrelated  HIV sequence; this PDF could be considered as the basis for searching sequences that are statistically negatively correlated to the HIV sequence, i.e.  the more similarity  between PDF of given sequence and this PDF, the more counteraction against HIV statistical treatment and probably its biological functionality \cite{wu2005statistical} }
				\label{PDF2}}
		\end{figure}
	\subsection{PDF estimation for the counteracting an original sequence }
	Suppose we take an original sequence of genome or a virus equivalent to a vector; in order to plausibly  estimate its empirical PDF, we are required to have access on an infinite ensemble of this vector 
	whereas, in our instance, such scenario is impractical, and even if we provide enough ensemble for a genomic sequence; i.e. random vector, next dilemma is the high-dimensionality issue of its PDF representation. For instance to estimate PDF of a virus such as HIV with the length 9181 can be represented in such	dimension. Therefore, {\em theoretical} PDF estimation strategy must be sought. In this regard, we propose Gaussian Mixture Model (GMM) \cite{zivkovic2004improved} to model the decimal sequence provided by Huffman coding for genome or a virus such as HIV sequence. Since this sequence and generally genomic sequences are highly correlated as seen in Fig.\ref{autocorr}, we estimate a two dimensional marginal PDF; as we see in Fig.\ref{PDF1}, using GMM of the original sequence of length $N$.
	\begin{equation}
		\label{eq02}
	p(x) = \sum\limits_{i = 1}^M {{w_i}} {\mkern 1mu} {\cal N}(x,{\mu _i},{\Sigma _i}),\quad 0<{w_i}<1
	\end{equation}
	where $p(\mbox{\boldmath $x$})=p(x_1,x_2)$ is the two dimensional marginal probability density function of a genomic sequence modeled by a GMM of order $M$ of Gaussian kernels, $\mathcal{N}(\mbox{\boldmath $x$},\mbox{\boldmath $\mu$}_i,\mbox{\boldmath $\Sigma$}_i)$, whose mean vectors and covariance matrices are denoted by $\mbox{\boldmath $\mu$}_i$ and $\mbox{\boldmath $\Sigma$}_i$, respectively. We assume the PDF for the vector $\mbox{\boldmath $x$}'=(x'_1,x'_2)$ counteracting the original sequence $\mbox{\boldmath $\x$}$ is also modeled by a GMM as follows
	\begin{equation}
		\label{eq03}
	p({x^\prime }) = \sum\limits_{i = 1}^M {{w_i}} {\mkern 1mu} {\cal N}({x^\prime },{\mu _i},{\Sigma ^\prime }_i),\quad 0<{w_i}<1
	\end{equation}
	that only differs in covariance matrices $\mbox{\boldmath $\Sigma$}'$ which	is related to $\mbox{\boldmath $\Sigma$}$ according to
	\begin{equation}
		\label{eq04}
		{\Sigma _{i{i^\prime }}} = {\Sigma _{ii}},\quad {\Sigma _{i{j^\prime }}} =  - {\Sigma _{ij}},\quad i \ne j,
	\end{equation}
	that is, the diagonal entries of both covariance matrices are the same but their off-diagonal entries are opposite in sign to rationalize the utmost negative statistical correlation tieing $\mbox{\boldmath $x$}$ and $\mbox{\boldmath $x$}'$. PDF of highly uncorrelated targeted sequence for HIV nocleotide sequence is also shown in Fig.\ref{PDF2}

	\section{Conclusion}
	In this paper, a new framework is presented in which, firstly we propose a preprocessing step to prepare 18 genomic sequences for further statistical modeling. Then  perform second step, statistical modeling, to capture linear and nonlinear characteristics of genomic sequence. Furthermore, we propose an approach to synthesize a new sequences based on our modeling for sequences of several original genes of the genome sequence and finally using GMM for PDF estimation, at first PDF of a given sequence such as HIV is estimated and based on that, another PDF is estimated which relates to the sequences with highly negative statistical correlation against the original sequence PDF, here HIV. This negative statistical correlation would suggest biological functional counteraction.
	%
\bibliographystyle{IEEEbib}
\bibliography{DNA,refs}

\end{document}